\documentclass[sigconf]{acmart}
\AtBeginDocument{%
  }

\copyrightyear{2025}
\acmYear{2025}
\setcopyright{cc}
\setcctype{by}
\acmConference[CHI '25]{CHI Conference on Human Factors in Computing Systems}{April 26-May 1, 2025}{Yokohama, Japan}
\acmBooktitle{CHI Conference on Human Factors in Computing Systems (CHI '25), April 26-May 1, 2025, Yokohama, Japan}\acmDOI{10.1145/3706598.3714254}
\acmISBN{979-8-4007-1394-1/25/04}

\usepackage{graphicx}
\graphicspath{{./fig/}}
\usepackage[acronym]{glossaries}
\usepackage{xspace}

\begin{document}

\title{VibWalk: Mapping Lower-limb Haptic Experiences of Everyday Walking}
\author{Shih Ying-Lei}
\email{23s151045@stu.hit.edu.cn}
\affiliation{
  \institution{HITSZ}
  \city{Shenzhen}
  \state{Guangdong}
  \country{China}
}

\author{Dongxu Tang}
\email{24s151056@stu.hit.edu.cn}
\affiliation{
  \institution{HITSZ}
  \city{Shenzhen}
  \state{Guangdong}
  \country{China}
}

\author{Weiming Hu}
\email{210110529@stu.hit.edu.cn}
\affiliation{%
  \institution{HITSZ}
  \city{Shenzhen}
  \state{Guangdong}
  \country{China}
}

\author{Sang Ho Yoon}
\email{sangho@kaist.ac.kr}
\affiliation{%
 \institution{KAIST}
 \city{Daejeon}
 \country{Republic of Korea}}

\author{Yitian Shao}
\email{shaoyitian@hit.edu.cn}
\affiliation{%
  \institution{HITSZ}
  \city{Shenzhen}
  \state{Guangdong}
  \country{China}}

\renewcommand{\shortauthors}{Shih et al.}

\newcommand{\oursystem}{\textit{VibWalk}\xspace}
\newcommand{\ShoePlate}{Vibration Transmitter Plate\xspace}
\newcommand{\shoeplate}{vibration transmitter plate\xspace}

\newcommand{\marked}[1]{{\textbf{\scriptsize\color[rgb]{1, 0.2, 0.3}{#1}}}}
\newcommand{\add}[1]{{\color[rgb]{0, 0.5, 0.1}{#1}}}
\newcommand{\modify}[1]{{\color[rgb]{0.0, 0.1, 0.6}{#1}}}
\newcommand{\delete}[1]{{\color[rgb]{0.8, 0.8, 0.8}{#1}}}
\newcommand{\um}[0]{\textmu m}
\newcommand{\sqrmm}[0]{mm\textsuperscript{2}~}
\newacronym{xr}{XR}{extended reality}
\newacronym{cnn}{CNN}{convolutional neural network}
\newacronym{tko}{TKO}{Teager-Kaiser operator}
\newacronym{anova}{ANOVA}{analysis of variance}

\begin{abstract}
Walking is among the most common human activities where the feet can gather rich tactile information from the ground. The dynamic contact between the feet and the ground generates vibration signals that can be sensed by the foot skin. While existing research focuses on foot pressure sensing and lower-limb interactions, methods of decoding tactile information from foot vibrations remain underexplored. Here, we propose a foot-equipped wearable system capable of recording wideband vibration signals during walking activities. By enabling location-based recording, our system generates maps of haptic data that encode information on ground materials, lower-limb activities, and road conditions. Its efficacy was demonstrated through studies involving 31 users walking over 18 different ground textures, achieving an overall identification accuracy exceeding 95\% (cross-user accuracy of 87\%). Our system allows pedestrians to map haptic information through their daily walking activities, which has potential applications in creating digitalized walking experiences and monitoring road conditions.
\end{abstract}  

\ccsdesc[500]{Human-centered computing~Mobile computing}

\keywords{Vibration sensing, wearable device, haptic information, haptic map}
\begin{teaserfigure}
  \includegraphics[width=\textwidth]{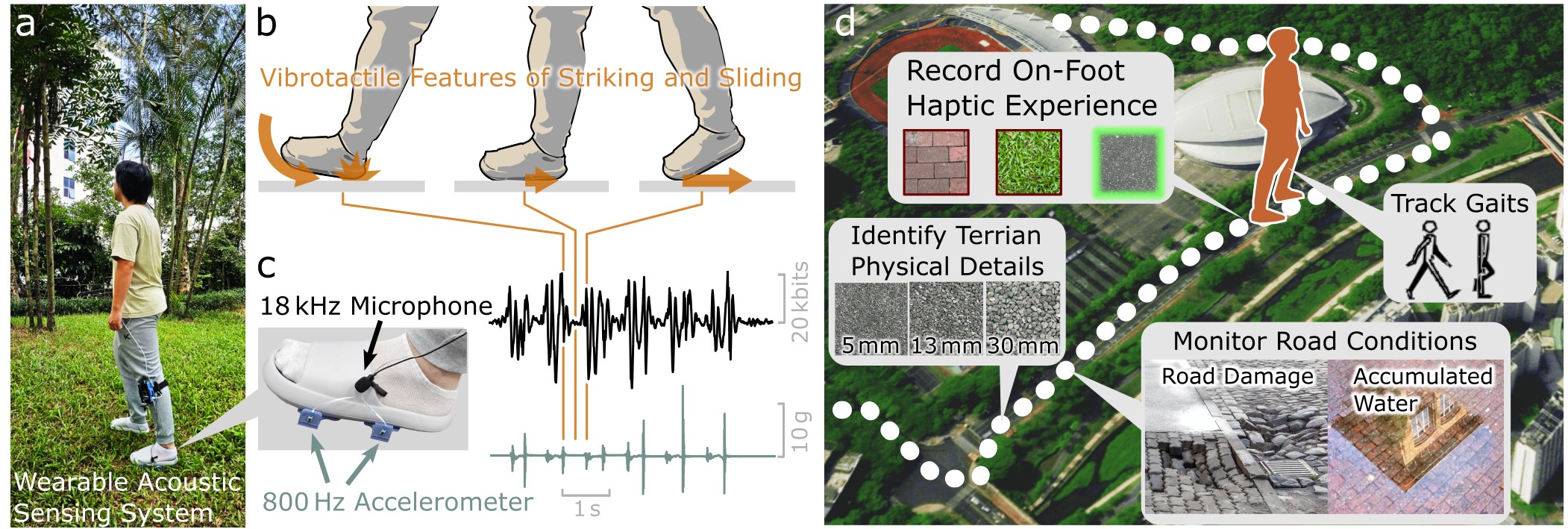}
  \caption{Capturing wideband vibration signals produced during natural walking activities. (a) A user wearing the wideband vibration sensing system. (b) A gait cycle contains striking and sliding interactions with the ground that produce vibrotactile features. (c) The MIC records the signals in the higher band (35$\sim$18000\,Hz), while the accelerometers capture the signals in the lower band (0$\sim$ 800\,Hz). (d) By enabling global positioning, \oursystem can capture and map the user’s haptic experience, including gait and on-foot vibrations, along with detailed road information throughout their walking path.}
  \Description{Capturing wideband vibration signals produced during natural walking activities. (a) A user wearing the wideband vibration sensing system. (b) A gait cycle contains striking and sliding interactions with the ground that produce vibrotactile features. (c) The MIC records the signals in the higher band (35$\sim$18000\,Hz), while the accelerometers capture the signals in the lower band (0$\sim$ 800\,Hz). (d) By enabling global positioning, \oursystem can capture and map the user’s haptic experience, including gait and on-foot vibrations, along with detailed road information throughout their walking path.
  }
  \label{fig:teaser}
\end{teaserfigure}

\received{20 February 2007}
\received[revised]{12 March 2009}
\received[accepted]{5 June 2009}

\maketitle
\section{Introduction}
The sense of touch enables humans to perceive their environment through body contact. For haptic research, tactile functions of the upper limb, especially the hand, have received increasing interest over the past few decades. In contrast, the tactile sensation experienced by the foot soles during walking has been less explored, despite the fact that the human foot is also among the most sensitive skin regions on the whole body. Human foot soles can sense dynamic skin deformation as small as 10\,\um~\cite{morioka2008vibrotactile} and gather a wealth of information during walking. A decline in the tactile sensitivity of the feet is considered one of the causes for increased risk of falls for elders~\cite{viseux2020sensory}, highlighting the importance of tactile information provided by the feet. However, the specific information encoded in the haptic signals generated by foot-ground contact during walking remains underexplored, particularly for signals spanning a wide frequency band.

Human bipedal walking utilizes repeated cycles of lower limb motion to move the body forward. Normally, the movement of a foot consists of the swing and stance phase determined by the state of contact between the foot and the ground surfaces \cite{pirker2017gait}. 
Existing works revealed the spatial information on the soles by using shoe-integrated measurement systems, including arrays of pressure or force sensors~\cite{paradiso2004interactive, matthies2017capsoles, ZULKIFLI202025}. 
On the other hand, when the foot makes contact with the ground, such as a heel strike, the entire footwear vibrates, and the foot skin can sense these vibrations with frequencies ranging from a few hertz to a kilohertz \cite{hayward2018brief}. By perceiving the vibrations, the pedestrian can identify the physical information of the ground. 
Studies indicate that the feet can discern ground compliance~\cite{visell2011vibration} and even perceive peripersonal space through perceived vibrotactile signals~\cite{amemiya2019remapping}. Thus, individuals can gather a significant amount of tactile information through their daily walking activities.

Long-distance walking enables pedestrians to perceive and map a vast region of terrains. Despite that haptic sensing provide aforementioned critical information for pedestrians, methods creating maps of haptic information has been less explored, as compared to vision-based mapping methods such as cameras \cite{davison2007monoslam} or lidars \cite{li20234d}.
Mapping the haptic information of pedestrians has many potential applications.
For instance, pedestrians can traverse narrow or uneven roads and assist urban management systems in monitoring poor road conditions, such as cracks or water accumulation. This is especially useful in situations where ground or air vehicles are disabled or prohibited \cite{alrajhi2023detection}.
Moreover, the mapped haptic data can serve as a training database for legged robots, enabling them to develop context awareness through haptic sensing on their feet~\cite{sojka2023learning}.
In addition, a map of haptic information can aid in developing lower-limb haptic feedback, enhancing the immersiveness of an \gls{xr} environment as users walk and explore the digital space \cite{visell2011walking}.
Thus, technologies capable of recording and mapping haptic information are desired.

In this paper, we present \oursystem, a foot-worn system designed to capture vibration signals generated during walking activities. Our findings demonstrate that these signals contain rich information, enabling the system to develop a haptic awareness of the surroundings, effectively creating a ``haptic map” (see in figure~\ref{fig:teaser}). \oursystem includes a device that can be mounted on a footwear to capture the vibration signals produced during walking activities. The sensing components contain accelerometers (frequency range 0$\sim$800\,Hz) that can record dynamic haptic signals produced by lower-limb movement and shoe-ground contact. We also used a microphone (frequency range 35$\sim$18000\,Hz) to record the haptic signals with high temporal resolution, which can facilitate applications that require a detailed spectral characterization or a precise reconstruction of signal waveforms.
Moreover, \oursystem can decode tactile information by utilizing lightweight machine learning architectures that can promptly extract temporal and spectral features of the vibration signals. We conducted studies involving 31 users walking over 18 different ground materials, achieving an overall material identification accuracy exceeding 95\% and cross-user identification accuracy over 87\%.
\oursystem can register the information on a map via geographical localization enabled by a GPS module. Thus, \oursystem can map the lower-limb haptic experiences of its users, including the vibrations felt by their feet, their walking gait, and physical details of the road surfaces.

The contribution of our work is summarized as follows:
\begin{itemize}
\item
A wearable system that enables a detailed map of haptic information utilizing pedestrian data analysis;
\item 
A haptic information mapping system that decodes types, textures, and conditions of the walking ground from vibration features and captures rich vibrotactile features during natural walking activities; and
\item 
A database of wideband vibrations was collected from 31 participants walking over 18 different ground surfaces, with six of these surfaces measured in both dry and wet conditions. Our database is available at: \href{https://huggingface.co/datasets/Tdongxu/Ground_material_classification/tree/main}{Ground Material Classification Dataset}
.
\end{itemize}

\section{Related Work}

\subsection{Footwear Sensing}
With advancements in wearable electronics, smart shoes can now monitor human activities for a wide range of applications~\cite{yan2022shoes++,saidani2018survey}, including navigation for the blind \cite{chava2021iot}, real-time monitoring of joint angles for the health of knee joint patients \cite{xia2017validation}. Some shoes are capable of monitoring the position of the foot inside can prevent falls of the elderly~\cite{ram2023novel}.  Smart shoes now can also monitor the health of the users, such as their weight~\cite{muzaffar2020shoe} or even diabetic foot ulcer~\cite{ghazi2024design}.

Different sensing principles have been employed in the design of smart shoes. One study explored the possibility of identifying ground materials using on-foot radar \cite{10.1145/3586183.3606738}. Another utilized laser speckle imaging for recognizing ground materials \cite{10.1145/3544548.3581344}. Human biosignals also contain rich information, and many sensing techniques have been developed. For instance, foot plantar pressure can be used to identify users \cite{matthies2017capsoles}. Smart insoles made of capacitive sensors can measure the pressure distribution on the sole to provide gait information \cite{luna2023smart}. Embedded force sensors can predict the ground reaction forces applied to the sole \cite{yamaguchi2023prediction} and perform gait analysis \cite{hamid2023development}. Inertial sensor measurements can also estimate human foot motion during normal walking \cite{yun2012estimation}.
Footwear sensing that utilizes vibrotactile signals, specifically the on-foot vibrations produced during walking, is yet to be explored.

\begin{figure*}[t]
  \centering
  \includegraphics[width=\linewidth]{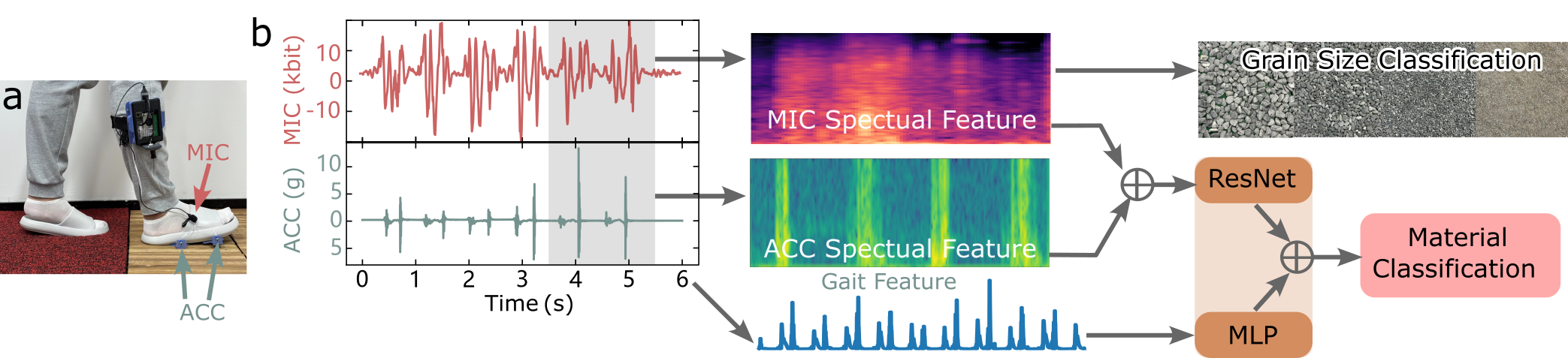}
  \caption{The system overview of \oursystem. (a) An example data collection scene including two ground materials: wooden board and carpet. \oursystem synchronously record signals from the two accelerometers (ACC) and the microphone (MIC) attached to the shoe. (b) The system extracts the spectral features and the low-frequency feature of the windowed vibration data to decode walking information, such as ground materials and grain sizes.}
  \Description{The system overview of \oursystem. (a) An examplary data collection scene including two ground materials: wooden board and carpet. \oursystem synchronously record signals from the two accelerometers (ACC) and the microphone (MIC) attached to the shoe. (b) The system extracts the spectral features and the low-frequency feature of the windowed vibration data to decode walking information, such as ground materials and grain sizes.}
  \label{fig:system}
\end{figure*}

\subsection{Vibration Sensing Techniques}
Vibrations, encompassing mechanical waves sensed via microphones from air and inertial measurement units (IMUs) from solid media, have been extensively used for recognizing user interactions and activities. IMUs capture mechanical vibrations through solids, aiding in human activity recognition (HAR). Ahmad and Leung \cite{ahmad2024hyperhar} proposed HyperHAR, learning bilateral correlations between sensing devices to improve wearable-based HAR. Hong et al. \cite{hong2024crosshar} introduced CrossHAR, enhancing cross-dataset HAR generalization via hierarchical self-supervised pretraining. Xu et al. \cite{xu2023practically} presented UniHAR, combining physics-guided data augmentation and self-supervised learning to address data heterogeneity across user groups. 
Contact elicited vibrations, such as surface acoustic waves (SAWs), can be used to detect user gestures on surfaces. Iravantchi et al. \cite{iravantchi2023sawsense} introduced SAWSense, leveraging SAWs for accurate surface-bound event recognition. Similarly, Lee et al. \cite{lee2024echowrist} developed EchoWrist, a wristband employing active acoustic sensing for continuous 3D hand pose tracking and hand-object interaction recognition with low power consumption.
Combining microphone and IMU data enhances recognition systems. Koch et al.~\cite{koch2024recognition} developed a multi-sensor system using ambient acoustic data for activity recognition in smart homes. Kunze and Lukowicz \cite{kunze2007symbolic} achieved symbolic object localization through the active sampling of acceleration and sound signatures with simple sensors.

Vibrations have been employed for environmental sensing. Concon et al.~\cite{concon2021deep} utilized IMU sensors for terrain surface classification, aiding robots in optimizing navigation. Microphone sensing was used by a hexapod robot to recognize ground materials \cite{christie2016acoustics}. Peng et al. \cite{peng2002vibration} improved vibration signal analysis for fault diagnosis using wavelet transforms. Yang et al. \cite{yang2015vibration} proposed a parameterized time-frequency method for feature extraction in varying-speed machinery, enhancing fault detection.

Deep learning models have enhanced performance in sensing complex environments. Laput et al. ~\cite{laput2018ubicoustics} presented Ubicoustics, a plug-and-play acoustic activity recognition system that achieves high accuracy without user training by utilizing professional sound effect libraries. Wu et al.~\cite{wu2020automated} proposed an incremental learning system for acoustic activity recognition, reducing user burden by prompting for labels only when necessary. Advanced deep learning architectures also improved multi-channel acoustic processing. Opochinsky et al. \cite{opochinsky2024single} introduced Sep-TFAnet, enhancing single-microphone speaker separation in noisy environments using attention mechanisms. Yang and Li \cite{yang2024self} proposed a self-supervised learning approach for spatial acoustic representation with cross-channel signal reconstruction, improving spatial acoustic parameter estimation. Wang et al. \cite{wang2023df} proposed DF-Sense, enabling multi-user heartbeat monitoring using acoustic sensing with dual-forming signal processing. Zhuang et al. \cite{zhuang2021reflectrack} introduced ReflecTrack, utilizing dual-microphone smartphones for 3D acoustic position tracking without additional hardware. Zhang and Wang \cite{zhang2022neural} presented a neural cascade architecture for multi-channel acoustic echo suppression. Still, utilizing deep learning to decode environmental information from haptic signals produced during natural walking remains unexplored.

\section{On-foot Vibration Sensing}
Pedestrians encounter a rich haptic experience during their daily walks.
Walking activity primarily consists of repeated cycles of lower limb motion, with the feet alternating between the swing and stance phases, marked by contact with the ground surface ~\cite{pirker2017gait}.
In stance phase, the contact area between the foot and the ground expands from the heel to encompass the entire sole, and then reduces until the toes lift off the ground. During this process, the foot perceives vibrations that provide detailed information about the ground surface. Specifically, the physical contact between the shoe and the ground produces vibrations that can spread from the contact locations to the entire shoe and sensed by the touch receptors inside the foot skin~\cite{shao2016spatial}.
By wearing wideband vibration sensors on the foot, this detailed information can be captured and utilized to identify the physical properties of the ground. An overview of our design is shown in Figure~\ref{fig:system}.

\begin{figure*}[h]
  \centering
  \includegraphics[width=\linewidth]{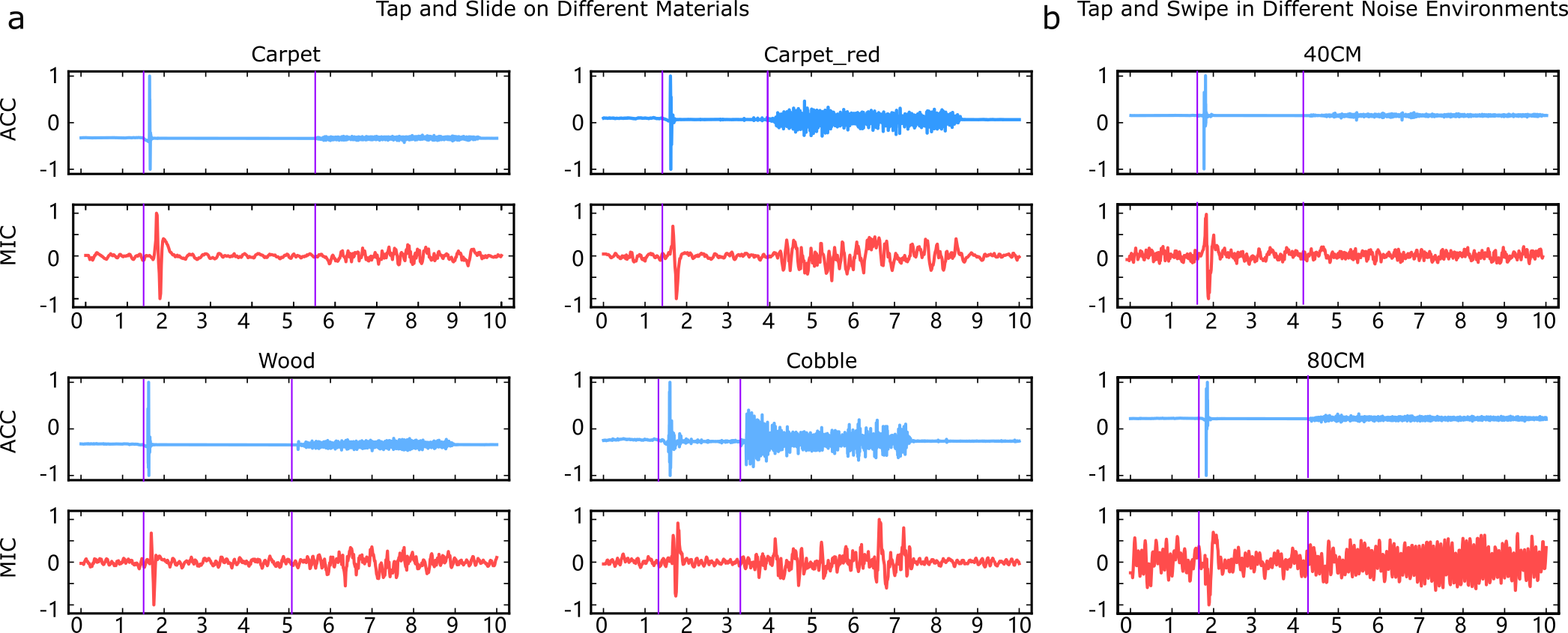}
  \caption{Assessment of ACC and MIC sensing. (a) A vibration transmission plate is used to tap and slide on four different materials. The first purple line on each picture represents the start of the tapping, and the second purple line represents the start of the sliding. (b) Tapping and sliding the plate under two different noise conditions. 40\,cm means a 60\,dB noise source is positioned 40\,cm from the sensors, and 80\,cm means the noise is 80\,cm away from the sensors.}
  \label{fig:validation}
\end{figure*}

\subsection{Sensing Principle}
Different materials exhibit unique responses to physical contact, producing vibrations with distinctive frequency characteristics \cite{kessler2002damage}, which is partially due to their difference in density, elastic modulus, and vibration attenuation factors \cite{sutherland2018review}.
Thus, vibration characterization has been used for a wide range of applications, from structural engineering to material analysis. Recent studies have highlighted the effectiveness of using vibration data, acquired with accelerometers and microphones, to characterize and identify materials \cite{khojasteh2024robust,christie2016acoustics,nie2023surface,strada2020leveraging}. 

In this work, we employed an accelerometer with a bandwidth of 800\,Hz to capture the vibrations produced from shoe-ground contact and to track the movement of the lower-limb. In addition, to capture vibration features in kilohertz levels, we incorporated a microphone with a sensing range between 35 to 18000\,Hz.
The combination of accelerator and microphone allows wideband sensing, including the low-frequency dynamics of lower-limb movement ($<10$\,Hz), the vibration features in the middle-frequency range ($10\sim400$\,Hz) that correspond to the most sensitive spectral range of human tactile sensing, and the high-frequency signals retaining the detailed waveforms of vibrations produced from shoe-ground contact.
For the rest of this paper, microphone sensing will be referred to as MIC, and accelerometer as ACC. However, environmental noise can impact the quality of vibrations in practical applications, and different sensors exhibit varying capabilities across frequency bands. To further explore the necessity of using both the ACC and MIC, we conducted two experiments to analyze the performance of these sensors in material feature extraction and noise robustness.

\subsection{Hardware Design}
\begin{figure*}[h]
  \centering
  \includegraphics[width=\linewidth]{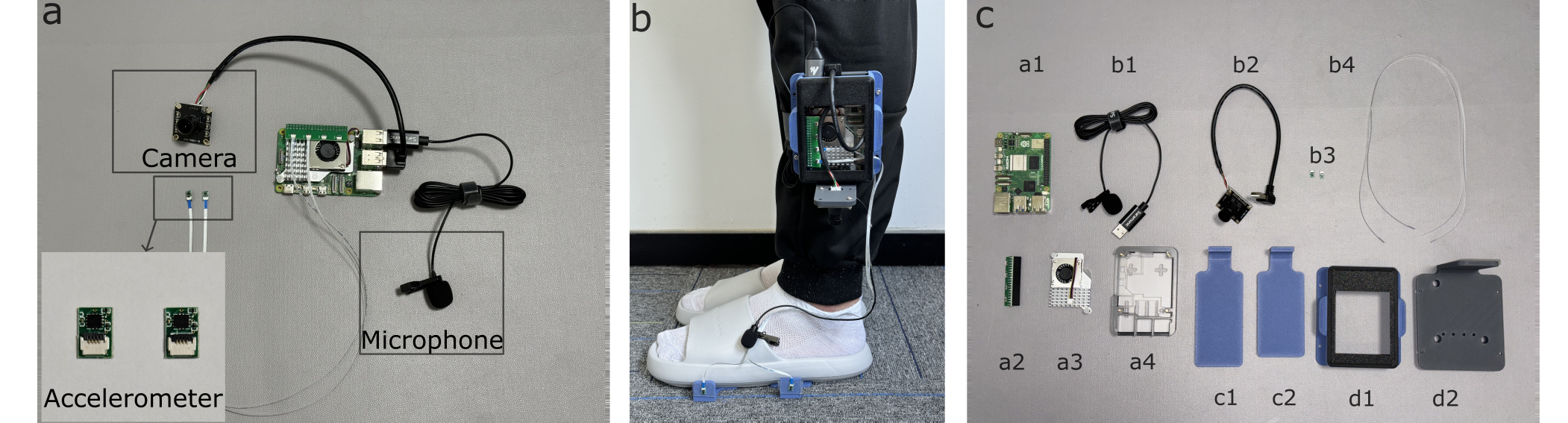}
  \caption{The hardware design of \oursystem.  
  (a) The fundamental sensing components of \oursystem; (b) \oursystem worn on a user; (c) Individual components, in which a1-a3 and b1-b4 are electronic components, c1-c2 are the mechanical components for affixing our accelerometers to shoes, and d1-d2 are mechanical components for housing the Raspberry Pi. a1-Raspberry Pi 5, a2-sensor contact plate, a3-radiator, a4-Raspberry Pi fixed shell, b1-microphone (MIC), b2-camera, b3-accelerometer (ACC), b4-sensor connection cable, c1-forefoot \shoeplate, c2-rearfoot \shoeplate, d1-square housing to cover the Raspberry Pi, d2-support component between a user’s lower leg and the hardware and carry the camera.}
  \Description{The hardware of \oursystem. (a)The circuit connection of electronic components; (b) Individual unit of the system,in which a1-a3 and b1-b4 are electronic components,c1-c2 are the mechanical components for affixing our accelerometers to shoes, and d1-d2 are mechanical components for housing the Raspberry Pi. a1-Raspberry Pi 5, a2-sensor contact plate, a3-radiator, a4-Raspberry Pi fixed shell,b1-audio sensor, b2-image sensor, b3-accelerometer, b4-sensor connection cable, c1-forefoot sensor fixing assembly, c2-rear foot sensor fixing assembly, d1-square housing to cover the Raspberry Pi, d2-support component between a user’s lower limb and the hardware and carry the image sensor.}
  \label{fig:fig3}
\end{figure*}

\subsubsection{Combining ACC and MIC for Wideband Vibration Sensing}
To demonstrate the unique sensing capability of the ACC and MIC on ground materials, we designed and 3D-printed a transmission plate using polylactic acid.
The transmission plate is equipped with an accelerometer and a microphone affixed to its surface. Experiments were conducted on four different material surfaces: wooden board, carpet, red carpet, and pebbles.
During the experiments, we first used the transmission plate to tap each material once, then waited for a short period before sliding the plate across the material's surface. The signals were simultaneously recorded by the ACC and MIC. The collected signals for each material are shown Fig.~\ref{fig:validation}a. It can be observed that both the ACC and MIC signals exhibit noticeable differences across the different materials. This indicates that both sensors are capable of capturing the distinctive vibration characteristics of materials, which is essential for material recognition and classification. 

\subsubsection{Noise Robustness Assessment} \label{section:noise-validation}
To assess the performance of the ACC and MIC under environmental noise, we designed a second experiment. In this experiment, a noise source emitting 60\,dB sound was placed at a distance of 80\,cm from the transmission plate. We then performed the tapping and sliding operations on the material. Subsequently, the noise source was moved to a distance of 40\,cm from the transmission plate, and the procedures were repeated.

The results demonstrate that, regardless of whether the noise source was placed at 80\,cm or 40\,cm, the sliding signals captured by the MIC were almost entirely masked by the environmental noise. In contrast, the ACC's sliding signals remained largely unaffected, clearly reflecting the vibrations generated by the sliding motion. This suggests that the ACC exhibits stronger robustness to environmental noise, effectively resisting interference in noisy conditions. We further validated the noise resistance capability of our system through a study in Section~\ref{section:audiowithnoise}.

The experiments validate the necessity of simultaneously utilizing both the accelerometer and the microphone. The MIC can capture high-frequency acoustic details up to 18\,kHz, ideal for recognizing fine material features, while the ACC demonstrates superior noise immunity, ensuring the reliability of the signals. By combining these two sensors, we achieve comprehensive sensing of materials and movements, enhancing the system's stability and accuracy in complex environments.

\oursystem can identify the surface of the ground by collecting the air-mediated audio signals and shoe-mediated vibration signals generated from dynamic physical contact between the foot and the ground. It consists of five parts: two customized accelerometers (STMicroelectronics AIS2IHTR, sensing range 0$\sim$800\,Hz), a miniature microphone (Saramonic SR-ULM10, sensing range 35$\sim$18000\,Hz), a miniature RGBD camera (Yahboo, 480P, 30\,FPS), a miniature computer (Raspberry Pi 5, with a Arm Cortex-A76 CPU), and customized 3D-printed assistive modules made from polylactic acid, as shown in Figure ~\ref{fig:fig3}b, including: module d1 and d2 tied to the legs to mount the Raspberry Pi; module c1 and c2 bonded to the bottom of the shoe. 
Here, modules c1 and c2 are vibration transmitter plates that can enhance the transmission of vibrations from the sole to the ACC sensors. Additionally, these plates can be easily attached to the bottom of shoes made from distinctive materials, while maintaining the overall generalizability of the sensing system.
The detailed hardware design of our system is shown in Figure ~\ref{fig:fig3}. 

Haptic sensing of \oursystem relies on ACC and MIC. The ACC is placed on the side of the shoe and attached to the \shoeplate using 3M double-sided tape. The MIC is placed on the tongue of the shoe such that it can capture air-mediated acoustic signals originating from shoe-ground contact. The horizontal positioning of \shoeplate on the shoe was determined based on the size of the shoe.

\section{Data-driven \oursystem Model for Decoding Ground Information} 
The task of detecting and identifying ground materials and conditions when walking on different surfaces is different from ground classification when stationary \cite{10.1145/3586183.3606738}. This is because people have different weights, gaits, and foot stepping forces; therefore, generalizing a classification model to a large group of users is challenging. Besides, the time duration of the foot contacting the ground and generating valid data is very short in a data segment. We used a 2-second segmenting window to ensure each data segment contains at least one signal pulse of foot-ground contact \cite{christie2016acoustics}.

\subsection{Ground Materials}
The choice of materials in our study was based on referencing prior research \cite{10.1145/3586183.3606738,10.1145/3544548.3581344}, we removed uncommon ground materials, such as steel plates. We selected materials that are common in urban life, such as tiles, asphalt, soil, and grass. 
In addition, we chose materials that belong to the same material category but provide nuance differences in vibration feelings when people walk on them, such as stones of different particle sizes.

\subsection{Experiment Setup and Data Collection}\label{section:18material}
\subsubsection{Ground surface materials}
\begin{figure*}[h]
  \centering
  \includegraphics[width=140mm]{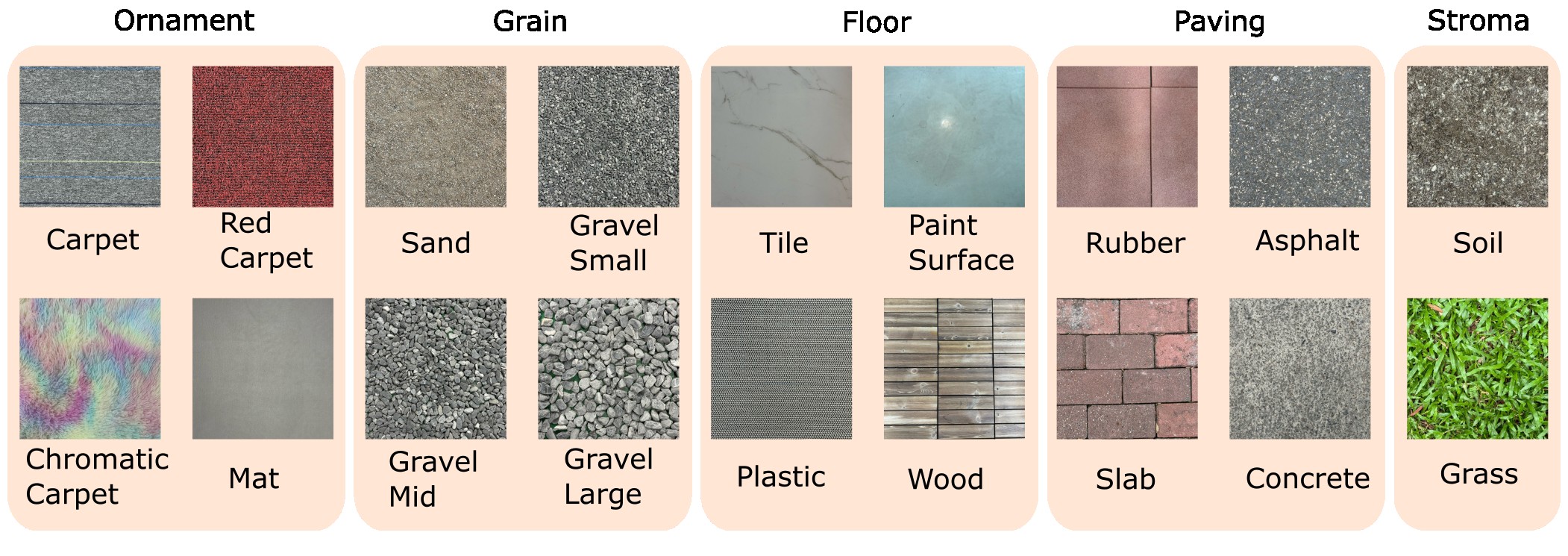}
  \caption{The photos of 18 ground materials used to study material classification. We divided the materials into 5 categories: Ornament, Grain, Floor, Paving, Stroma.}
  \Description{This is a diagram of 18 surface materials used to study material classification. We divided the 18 grounds into 5 categories, which are: Ornament, Grain, Floor, Paving, Stroma}
  \label{fig:18surface}
\end{figure*}

We studied a total of 18 types of ground materials, comprising 5 outdoor and 13 indoor materials. These materials are categorized into five groups: Ornament, Grain, Floor, Paving, and Stroma. Ornament is used for decorative floors in homes. Grain has different grave sizes. The floor is the surface used for indoor use. Paving is often used for road paving in urban construction. Stroma is the ground that already exists in nature and has not been artificially modified.
The photos of these ground materials are shown in Figure ~\ref{fig:18surface}.

\subsubsection{Participants and apparatus}
We recruited 31 participants (21 males and 10 females) ranging in age from 21 to 30 years old (mean = 23.48, SD = 1.56).
Their height ranged from 155\,cm to 187\,cm (mean = 172.6, SD = 7.4), and their weight ranged from 48\,kg to 90\,kg (mean = 66.7, SD = 11.6). In addition, all participants recruited in this experiment received their consent and signed the informed consent form. 
Our evaluation study was approved by the IRB at the author's institution.

\subsubsection{Data collection procedure}
We used the same data collection process for each subject. We had the subjects walk at a natural gait on 18 different ground materials. Each subject walked on each material for 100\,s. During this time, we collected ACC and MIC data on the subjects' interaction with the ground while walking. Of the 18 materials, 13 were laid inside buildings in a quiet environment, allowing us to eliminate street noise during the experiment. The remaining 5 materials are outdoor materials, and the MIC data were affected by external noise, such as vehicle noise, that contaminated the MIC measurements (See our investigation in Section~\ref{section:audiowithnoise}). Since we have one microphone, we collected data for 55,800\,s, or 15.5\,h, for 31 people and 18 materials for each person. And we have two accelerometers, so the accelerometer data was collected for a total of 111,600\,s, or 31\,h.

\subsection{Data Pre-processing}
Our model extracts both temporal and spectral features of vibrations produced from contacting surfaces to facilitate the recognition of ground materials \cite{cheng2022machine}. 

\paragraph{ACC features}
Our vibration sensing system was installed on only one of the user’s feet. The vibrations, originating from the foot-ground contact, travel through the shoe and are captured by the ACC sensors attached to that shoe. We set the window length to two seconds, and in order to prevent the data in the training set from overlapping with the data in the test set, we set the window sliding length to 2\,s and the stride to 2\,s, that is, there is no overlap between any two data segments. For 111,600 seconds of data, we divided it into 55,800 data segments. When extracting ACC features,the conversion of Mel spectrum to decibel data would enhance low-frequency data, while the weight of high-frequency data would be greatly reduced. In acceleration data, most low-frequency data is gait data of people walking, while high-frequency data of interaction with the ground would be mostly ignored or reduced. Therefore, we do not use Mel spectrum, but use short-time Fourier transform (STFT) with a more even frequency distribution. We set the length of each STFT window to 50 samples and the overlap of each window to 75\%.
We implemented high-pass filtering with a cut-off frequency of 20\,Hz to reduce the impact of lower-limb movement on our ACC sensing for the material recognition task. 

In addition, gait information can be extracted from the accelerometer data. By incorporating gait-related features into the model, the accuracy of material recognition can be improved. We adopted a lightweight feature extraction algorithm called \gls{tko} \cite{flood2019gait}, considering both the possibility of real-time detection brought by the lightweight model and ensuring the reliability of high-precision recognition processing. 
To extract \gls{tko} features, the ACC signals are filtered to 20\,HZ, and then used to calculate the \gls{tko} energy operator:
\begin{equation}
\phi_{n} = \frac{2x_{n}^{2} + (x_{n+1} - x_{n-1})^{2} - x_{n}(x_{n+2} + x_{n-2})}{4T_{s}^{2}}
\end{equation}

Where $\phi_{n}$ is the \gls{tko} energy operator, $x_{n}$ is the acceleration data after the offset of two samples, and $T_s$ is the sampling period. In this process, we did not perform the standard half-wave rectification, as we found that the signal without half-wave rectification was more stable for the model. The peak values were smoothed using a combination of a sample moving maximum window and a sample moving average window:
\begin{equation}
\phi'_{n} = \frac{1}{5} \sum_{n-2}^{n+2} \max(\phi_{i-1}, \phi_{i}, \phi_{i+1})
\end{equation}
The \gls{tko} feature vector $\phi'$ extracted from the ACC signal continues to retain the length of 3200, given as an input to the \oursystem model.

\paragraph{MIC features}
Unlike the ACC sensing, although we put only a single MIC sensor on one shoe, the MIC can still pick up the sound of the other shoe touching the ground. Therefore, we use a one-second window to ensure that a data segment has a valid signal of contact with the ground. For 55,800 seconds of data, we divided it into 55,800 data segments. We set the number of Mel filter banks to 64 and the number of samples between two consecutive frames to 800. 
We also high-pass filtered the MIC signals with a cut-off frequency of 20\,Hz to eliminate voltage drift issues in the analog-to-digital conversion module of the MIC sensor. This approach proved to be beneficial for enhancing classification performance.

\paragraph{Fusion features}

In order to facilitate the fusion of the two modalities, we converted the data of ACC and MIC into Mel spectrum. We set the number of Mel filter banks in MIC to 64 and the number of samples between two consecutive frames to 800 and in ACC we set the number of Mel filter banks in MIC to 64 and the number of samples between two consecutive frames to 80. Then we fuse them according to the dimensions of the Mel filter banks to form a 64 * 102 spliced Mel spectrum.

\subsection{\oursystem Model}
\begin{figure*}[h]
  \centering
  \includegraphics[width=160mm]{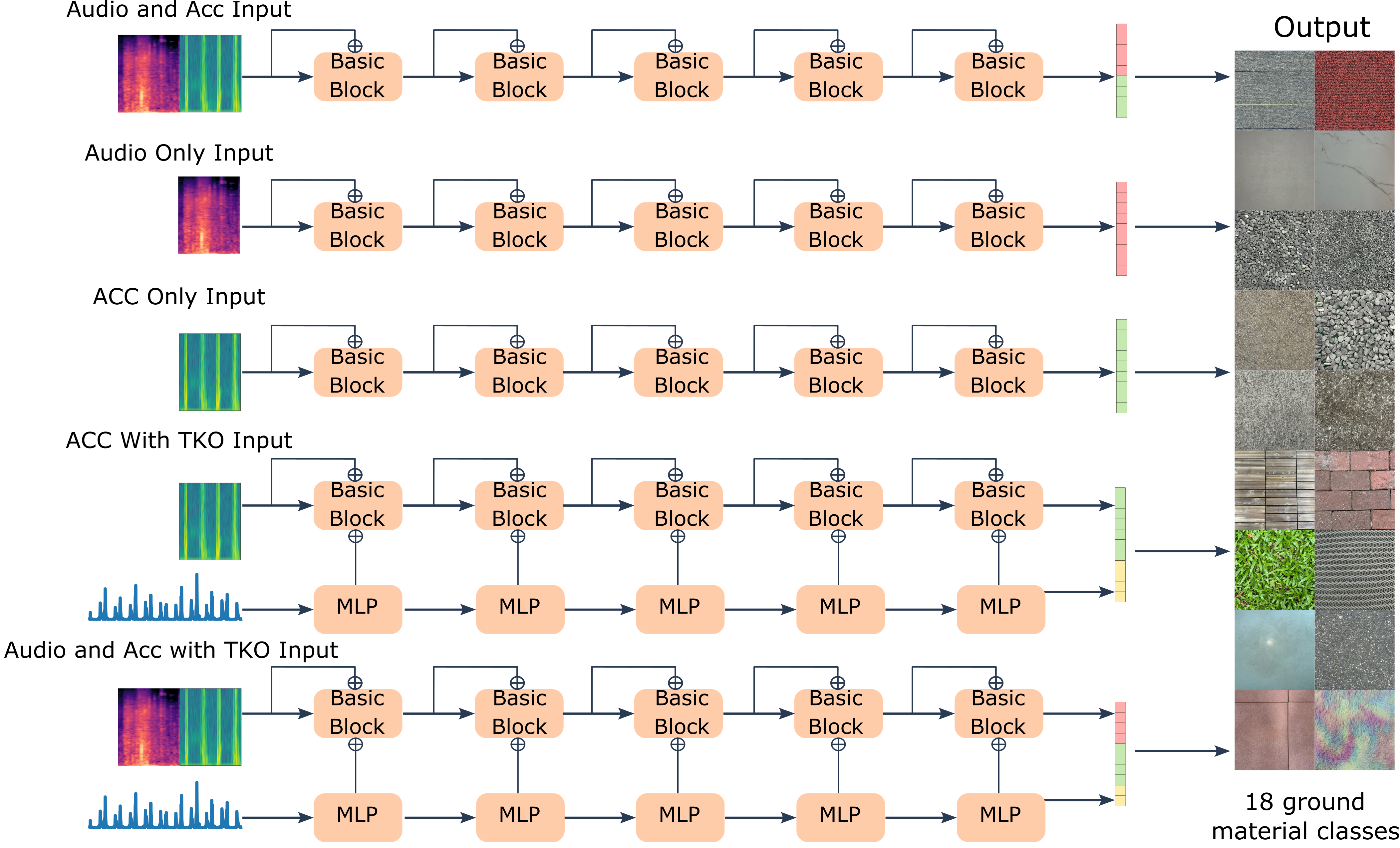}
  \caption{Our \oursystem model was designed based on a ResNet framework. We have a total of five \texttt{BasicBlock}, each with two convolutional layers. The numbers of channels in those blocks are 32, 64, 128, 256, and 512, respectively. In addition, for the MLP(Multi Layer Perceptron) layer, the first layer has an input channel of 3200 and an output channel of 128, and all subsequent layers have an input channel and an output channel of 128. \texttt{BasicBlock} comprises the following key components: Convolutional and Normalization Layers, Residual Connection (Shortcut), Auxiliary Input Processing (Optional), Forward Propagation}
  \Description{Our \oursystem model was designed based on a ResNet framework. We have a total of five basic residual blocks, each with two convolutional layers. The numbers of channels in those blocks are 32, 64, 128, 256 and 512, respectively. In addition, for the MLP layer, the first layer has an input channel of 3200 and an output channel of 128, and all subsequent layers have an input channel and an output channel of 128.}
  \label{fig:network}
\end{figure*}

We compared classical machine learning models and deep learning models and chose the one with the best performance. Their performance was evaluated based on the training time, the prediction time, and classification accuracy.
When comparing the data training time, we found that due to the large amount of data and the high dimensionality, the time for machine learning based on CPU computing has no advantage over deep learning based on GPU computing. The performance is presented in Table ~\ref{tab:performance}. The running times of different models on our data are shown in Table 1. 

\begin{table}[]
\caption{Training time and accuracy of different machine learning models on our data containing 18 ground materials.}
\begin{tabular}{@{}c|ccc@{}}
\toprule
Model                   & ResNet & MobileNetV3 & SVM \\ \midrule
f1-score($\%$)          & 94.8  & 89.0       & 78.0  \\
Time/train-loop(minute) & 66     & 54          & 56  \\ \bottomrule
\end{tabular}
\label{tab:performance}
\end{table} 

We devised \gls{cnn} models that take spectrograms as their input, a design commonly used in classification tasks for vibration data \cite{palanisamy2020rethinking}. 
Specifically, spectral features were extracted from time series signals so that the \gls{cnn} model can learn both the temporal and spectral patterns of the input data \cite{DBLP:conf/aaai/LiuLK0P24}.
Considering the limit of deploying deep learning models on a mobile device, We tested two lightweight \gls{cnn} frameworks, ResNet \cite{he2016deep} and MobileNetV3 \cite{koonce2021mobilenetv3}, together with a classical machine learning model, support vector machine \cite{hearst1998support}. The results suggest that ResNet excels in classfication accuracy, with only a minor increase in training time (Table~\ref{tab:performance}).

Our design of the \oursystem model adapted the ResNet framework. Because the dimension of our spectrogram is different from the 224*224 dimension commonly used in image processing, we modified the input of the original ResNet model to adapt to the dimension of our data. 

In this study, we utilize a fundamental residual module, referred to as \texttt{BasicBlock}, as the foundational building unit of our model. The design of this module is inspired by the residual network (ResNet) architecture, aiming to effectively mitigate the vanishing gradient problem in deep neural networks and facilitate the flow of information and feature propagation.

Specifically, the \texttt{BasicBlock} comprises the following key components:

\begin{enumerate}
    \item \textbf{Convolutional and Normalization Layers}:
    The module begins with two $3 \times 3$ convolutional layers, denoted as \texttt{conv1} and \texttt{conv2}. Each convolutional layer is followed by a two-dimensional batch normalization layer (\texttt{BatchNorm2d}). The stride of the first convolutional layer can be adjusted as needed to control the spatial dimensions of the feature maps. A ReLU activation function is applied to introduce non-linearity, enhancing the model's expressive power.
    
    \item \textbf{Residual Connection (Shortcut)}:
    To facilitate residual learning, the module includes a direct shortcut connection. When there is a mismatch in the number of input and output channels or when the stride is not equal to one, the shortcut connection is adjusted using a $1 \times 1$ convolutional layer followed by a batch normalization layer to match the dimensions of the main path. Otherwise, the input is directly incorporated into the shortcut connection without modification.
    
    \item \textbf{Auxiliary Input Processing (Optional)}:
    The module supports an optional auxiliary input $y$, which allows the incorporation of additional information. When the \texttt{use\_y} flag is set to true and \texttt{y\_dim} is specified, the auxiliary input $y$ is processed through a fully connected layer (\texttt{Linear}) and a layer normalization layer (\texttt{LayerNorm}). This projects $y$ to the same dimensionality as the residual connection, enabling the fusion of external information with the main path's output. This mechanism enhances the model's ability to perceive and integrate external information, thereby improving overall expressive capability.
    
    \item \textbf{Forward Propagation}:
    During forward propagation, the input $x$ first passes through the main path comprising convolutional and normalization layers, followed by ReLU activation. The output is then combined with the shortcut connection. If an auxiliary input $y$ is utilized, its processed form is also added to the combined output. Finally, a ReLU activation function is applied to produce the fused feature representation.
\end{enumerate}

\begin{table}[]
\caption{The time it takes to train ResNet models with different numbers of layers, given MIC data with two-second segmentation. The model was deployed on a computer with only a single CPU. ResNet models with different sizes (number of parameters) were explored.}
\begin{tabular}{@{}c|ccc@{}}
\toprule
ResNet Layers  & Prediction Time & F1 Score($\%$)    & Parameters\\ \midrule
6       & 16.45ms        & 92.38              & 888114            \\
8       & 19.21ms        & 94.25              & 1513010            \\
10 (\oursystem)     & 8.97ms         & 93.91              & 2539314\\
12       & 8.32ms  & 94.46              & 6177074\\
18      &11.53ms         & 94.62              & 21206706\\ \bottomrule
\end{tabular}
\label{tab:layers}
\end{table}

The \texttt{BasicBlock} module effectively integrates the main path with the shortcut connection and optionally incorporates auxiliary input processing. This combination facilitates efficient feature extraction and information fusion, thereby enhancing the model's expressive power and training stability.

We investigate the optimal configuration of ResNet for the application of material recognition (Table~\ref{tab:layers}). The final design of our network structure is shown in Figure ~\ref{fig:network}. The output of the adapted ResNet model passes through a fully connected layer to predict the labels of the ground materials.

As verified in section~\ref{section:noise-validation}, MIC and ACC signals have different advantages in different situations. Therefore, we need to train models for each modality to ensure that we can use the appropriate modality in specific situations, which requires using models with different parameters for different signals. We keep the convolution part of ResNet unchanged and only change the first fully connected layer to adapt to different signal inputs.

While training time increases with the number of layers, the execution time of a single data sample on the CPU decreases (Table~\ref{tab:layers}). Since downsampling is performed every two layers (that is, the height and width of the feature are divided by two), the 6- and 8-layer networks enter the flatten layer without sufficient downsampling, resulting in a longer CPU running time. This reduction in run-time is advantageous for establishing a real-time system. Given that the increase in training time is relatively modest, we ultimately selected the 10-layer ResNet as our primary model for both training and testing.

\section{Technical Evaluation of \oursystem Model}

\subsection{Results}
\begin{figure*}[h]
  \centering
  \includegraphics[width=160mm]{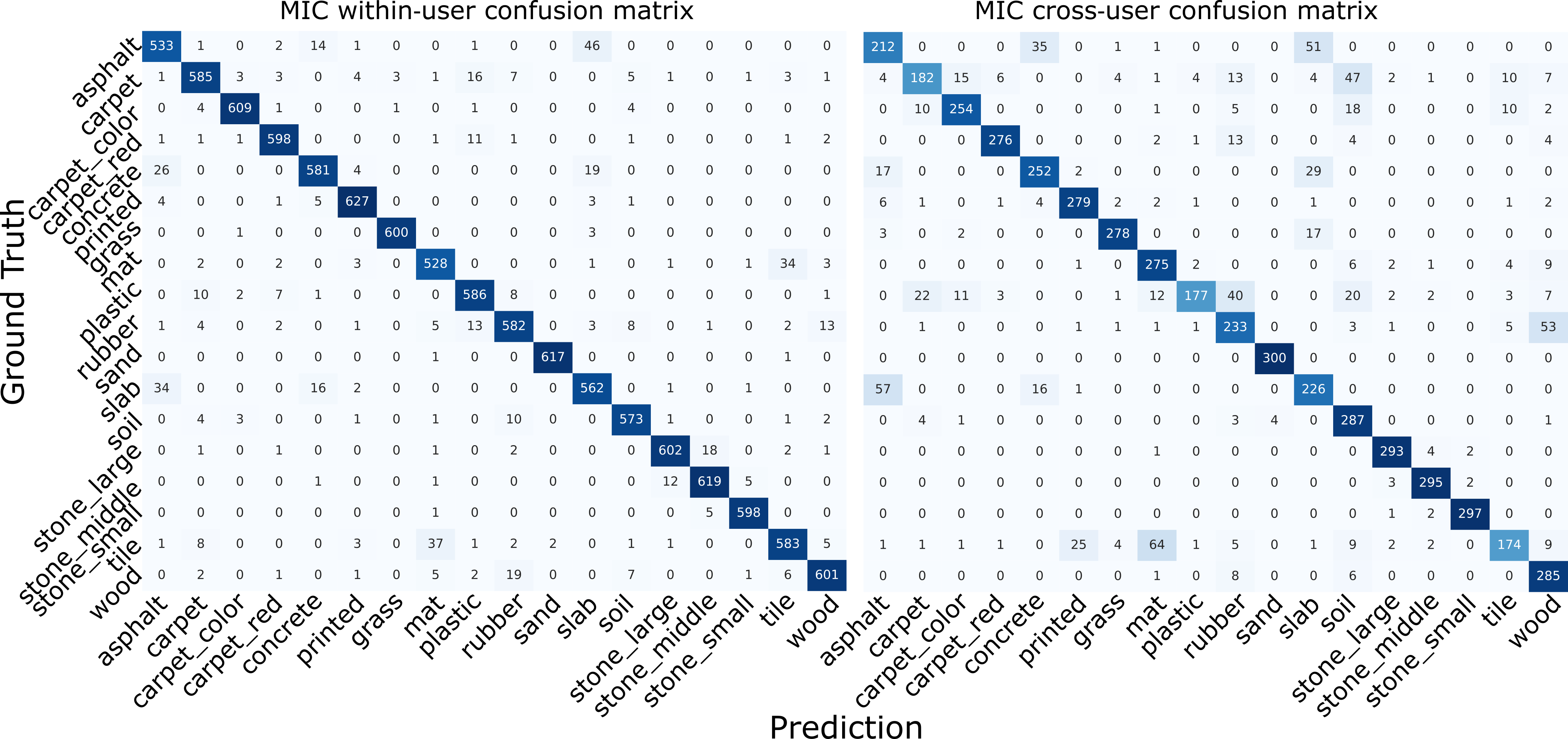}
  \caption{The confusion matrix of the MIC data test. Left: within-user; right: cross-user.}
  \Description{}
  \label{fig:MICCON}
\end{figure*}

\begin{figure*}[h]
  \centering
  \includegraphics[width=160mm]{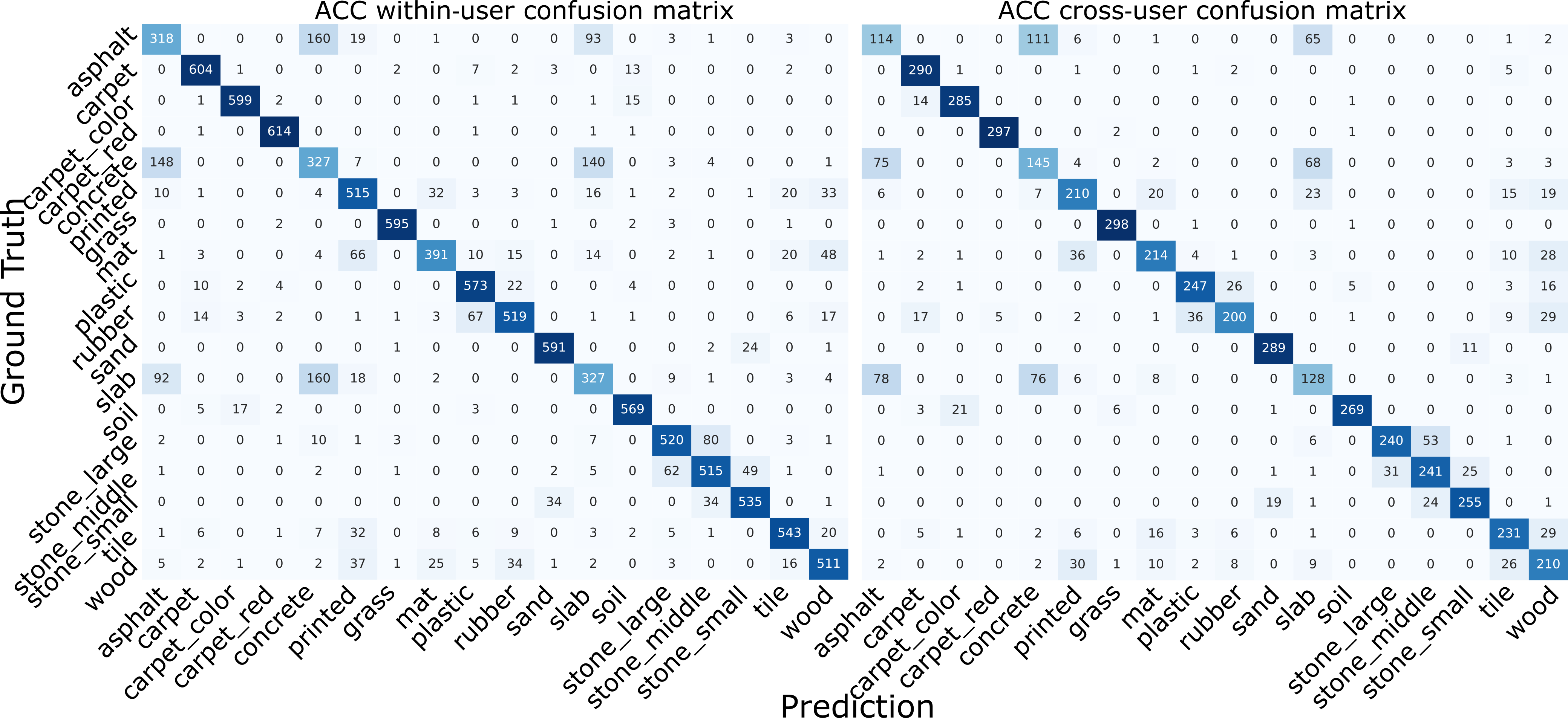}
  \caption{The confusion matrix of the ACC data test. Left: within-user; right: cross-user.}
  \Description{}
  \label{fig:ACCCON}
\end{figure*}

In order to evaluate the performance of two different sensor signals in ground material analysis, we used two different evaluation methods: within-user and cross-user. Among them, our within-user experiment used 10-fold cross validation, that is, our total 55,800 data samples were randomly divided into 10 parts. For each fold validation, nine parts of the data were selected as the training set and the remaining part was used as the validation set. In the cross-user evaluation, we divided the 31 subjects into nine groups of three and one group of four, for a total of ten groups. For each fold validation, we used nine of the groups as training sets and the remaining one as validation sets. Another point to note is that, whether it is within-user or cross-user experiments, the data of each material is close to being balanced. Therefore, it is possible to use accuracy and f1-score as the final evaluation criteria, but for safety reasons, we finally chose the more reasonable f1-score as our evaluation criterion. We initially assessed the sensing capabilities of the MIC and ACC independently, followed by their combined performance. Additionally, our model leverages gait information by extracting the \gls{tko} feature from the ACC data. The results are summarized in Table~\ref{tab:Ablation}.

\begin{table*}[]
\caption{F1 scores of \oursystem model with different input modalities, includes ACC only, MIC only, and the fusion of ACC and MIC data, ACC with \gls{tko}, the fusion of ACC, MIC and TKO data.}
\begin{tabular}{@{}c|ccccc@{}}
\toprule
            & MIC   & ACC & MIC+ACC &ACC (TKO) &MIC+ACC (TKO)\\ \midrule
Within-user & 92.9\% & 81.2\%      & 95.4\%   &84.2\%   &96.1\%\\
Cross-user  & 85.9\% & 70.6\%      & 87.1\%    &73.4\%  &88.5\%\\ \bottomrule
\end{tabular}
\label{tab:Ablation}
\end{table*}

\subsubsection{Evaluations based on MIC data}
The result of ground material classification based solely on MIC data is shown in Figure~\ref{fig:MICCON}, with within-user and cross-user evaluation.
Among them, the average f1-score of the ten-fold cross-validation of within-user is 92.9$\%$~(SD = 0.3). The confusion matrix exhibits a diagonal pattern. For cross-user evaluation, the average f1-score of the ten-fold cross-validation is 85.9$\%$~(SD = 2.6), around a 7.0\% drop compared to the within-user case. The accuracy dropped as a few groups of materials are less separable, such as carpet and soil, rubber and wood, tile and mat.

\subsubsection{Evaluations based on ACC data}
The result of ground material classification based solely on ACC data is shown in Figure~\ref{fig:ACCCON}, with within-user and cross-user evaluation included.
Among them, the average f1-score of the ten-fold cross-validation of the within-user case is 81.2$\%$, and the average f1-score of the ten-fold cross-validation of the cross-user case is only 70.6$\%$. The ACC confusion matrix reveals notable confusion between asphalt, slab, and concrete (Figure ~\ref{fig:ACCCON} and Section~\ref{appendix:acc18and16}), which underscores the importance of integrating additional sensing methods, including the MIC. With \gls{tko} feature extracted and used for classification task, the within-user case is increased to 84.2\%, the cross-user case is increased to 73.4\%. 

\subsubsection{Evaluations based on fusion data}

In this experiment, we fused the features of the ACC and MIC data before putting them into the deep model and extracted their Mel spectrum. The feature parameters here are slightly different from the parameter settings in Section~\ref{section:18material}. The MIC parameter settings of our spectrum data remain unchanged from the main experiment, and the final feature dimension is still 64*61. The ACC data is changed from the original STFT to Mel spectrum. We set n-mels to 64 to match the audio data, and finally generated a feature dimension of 64*41.
The average f1-score of the ten-fold cross-validation of Within-user is 95.4$\%$ (SD = 0.3), around 2.5\% increase in classification accuracy as compared to the MIC-only model and 14.2\% increase compared to the ACC-only model. In addition, we also put the \gls{tko} operator into it for training and testing. In the ten-fold test, the accuracy of within-user was 96.1\% and cross-user was 88.5\%, which was an improvement compared to without \gls{tko}.

\subsection{Analysis}
The classification results suggested that distinguishing between certain materials (such as concrete, slab, and asphalt) using ACC signals alone proved challenging.  
In the later stages of the study, we combined these three materials into a single category and retested the classification. The results showed that this adjustment improved the cross-user analysis accuracy from 0.70 to 0.78. This suggests that when different materials are difficult to distinguish at the signal level, appropriate category merging can reduce classification errors and improve the overall performance of the model. The successful application of this strategy not only overcame the classification difficulties caused by signal similarity but also demonstrated the feasibility of improving model accuracy by adjusting classification standards in complex environments. This finding provides valuable insights for future research, particularly in optimizing classification methods when dealing with highly similar data features to maximize classifier efficiency.

In terms of data form, some studies have utilized advanced sensors such as radar and laser imaging, which offer unique capabilities for ground material classification. Radar technology can probe deep-level ground, including composite grounds, which is challenging with traditional sensors. Laser imaging, on the other hand, functions as an image sensor, employing image processing techniques for data preprocessing, providing intuitive image data and enabling specific experiments unique to this type of sensor. Our research, however, focuses on the application of vibrations to analyze the interaction between people and the ground. With MIC and ACC sensing, we explore the interaction between individuals and the ground, going beyond mere ground material identification. Our approach allows us to capture a broad range of walking features and provides a deeper understanding of how different surfaces affect the haptic experience of walking.

\section{Enhancement of \oursystem Model}
\subsection{Audio with noise}\label{section:audiowithnoise}
In the study exploring the effect of environmental noise on two different sensing modalities, ACC and MIC, we recruited five subjects and played the same white noise sound in the experiment room via a loudspeaker. The data collection and processing procedures are the same as those in Section \ref{section:18material} except we collect the data with noise. The same noise is played when collecting vibration data for each material. We use thirteen materials in total: Carpet, Red Carpet, Chromatic Carpet, Mat, Sand, Gravel Small, Gravel Mid, Gravel Large, Tile, Plastic, Wood, Rubber, and Soil.

For evaluation, we used the same model trained based on the data collected in Section~\ref{section:18material} and tested it using the newly collected data contaminated with environmental noises.
As expected, the MIC test accuracy dropped significantly to 9.0$\%$, suggesting a heavy influence of environmental noises over the MIC sensing modality. Surprisingly, the classification accuracy of the ACC data also dropped, from 70.6\% to 62.0\%. Still, the results suggest that the performance of ACC sensing is much more robust compared to the MIC.

\subsection{The Role of the \ShoePlate}
When two materials collide, the generated signal is related to both materials \cite{sutherland2018review}. The design of \oursystem is intended to capture the intrinsic vibration features of the ground materials; thus, we controlled the material of the sole region that collides with the ground by placing a \shoeplate at the bottom of the shoe and mounted the ACC sensors directly on it. Such a design ensures that the vibrations are generated by the same ground-contact material attached to the bottom of the shoe, regardless of the different shoes worn by users of \oursystem.

We validated our design by collecting vibration data twice from the same subject wearing the laboratory experiment shoes, with and without the \shoeplate. In this study, we used all the materials. We had one participant perform data collection twice, the first time wearing sneakers without \shoeplate and the second time wearing sneakers with \shoeplate.
Without the \shoeplate, the classification accuracy of ground materials was 88.7\% for MIC and 73.7\% for ACC. Once the \shoeplate is mounted, the classification accuracy rose to 97.6\% for MIC and 80.6\% for ACC.

\section{Decoding Ground Detail}
In addition to material recognition, \oursystem was designed to capture detailed information about ground surfaces, such as water accumulation (Section~\ref{section:drywet}). A wetted ground surface exhibits distinctive vibration features compared to a dry one. Moreover, we show that the spatial characteristics of ground surfaces can be decoded simply based on the spectral features of the vibrations (Section~\ref{section:grainsizes}).

\begin{figure*}[h]
  \centering
  \includegraphics[width=160mm]{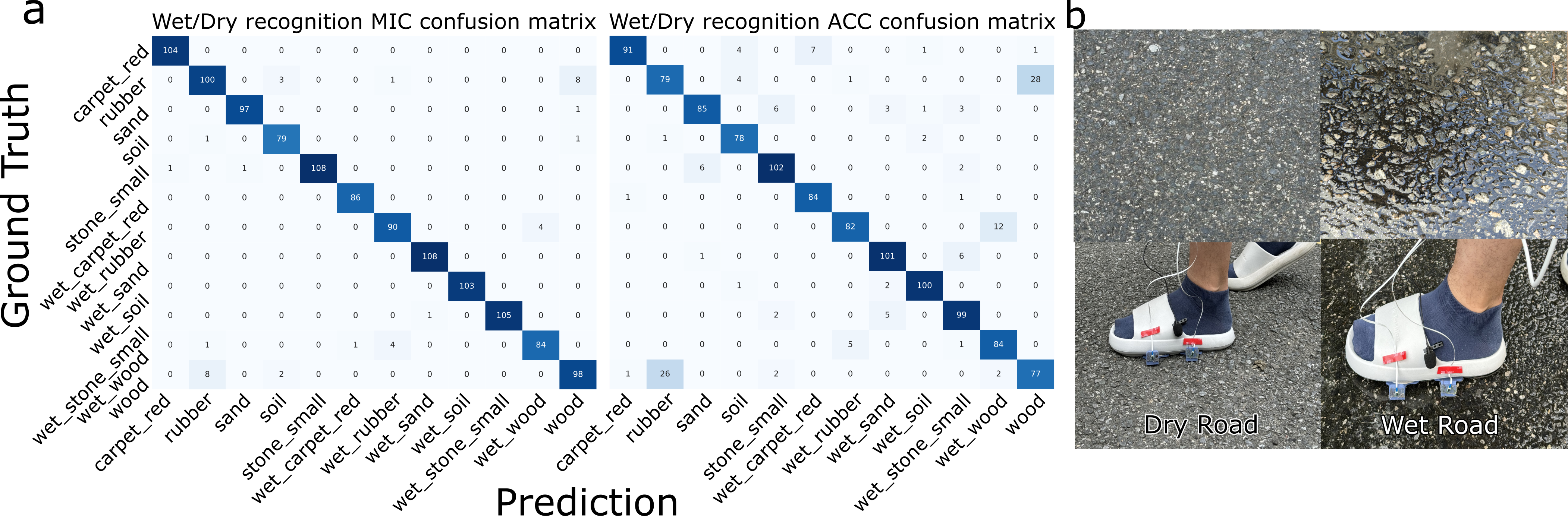}
  \caption{(a) The confusion matrix resulted from the Wet/Dry experiment using MIC (left) or ACC (right) sensing. (b) Examples of a dry (left column) and a wet (right column) ground.}
  \Description{}
  \label{fig:wetdry}
\end{figure*}

\subsection{Dry and Wet Surface}\label{section:drywet}
\subsubsection{Evaluation}
In the study of dry and wet floors, we recruited five participants. We selected six representative ground materials. These materials are commonly found in outdoor environments and often become wet during rainy weather. Accumulated water can erode these materials or make them hazardous for pedestrians.
These six materials are soil, rubber, sand, gravel, wood, and carpet. In this experiment, we found five subjects, and these five subjects collected data in both dry and wet conditions. The data collection and processing procedures are the same as those in Section \ref{section:18material}. We asked the subjects to collect data for 100 seconds on both dry and wet ground. For each subject, a total of 1,200 seconds were collected, and a total of 6,000 seconds were collected for the five subjects. We divided the data into 80$\%$ training set and 20$\%$ test set. The final results of MIC and ACC data are shown in Figure ~\ref{fig:wetdry}. The f1-score of MIC reached 96.9$\%$, while the f1-score of ACC reached 90.1$\%$.

\subsubsection{Application}
\oursystem has the capability to detect urban water accumulation. 
When users walk on a dry surface, the signals captured by the vibration sensors differ significantly from those produced from walking on a waterlogged surface. The presence of water alters the vibration patterns and feedback characteristics of the ground, resulting in specific signal changes. \oursystem can detects these changes and identify water accumulation, providing real-time reports on the water accumulation problem. This functionality is particularly important for urban environmental management, as it helps relevant authorities address water accumulation issues promptly, reducing potential safety hazards and improving the management efficiency of urban infrastructures.

\subsection{Identification of Grain Size}\label{section:grainsizes}
\subsubsection{Evaluation}
\begin{figure*}[h]
  \centering
  \includegraphics[width=160mm]{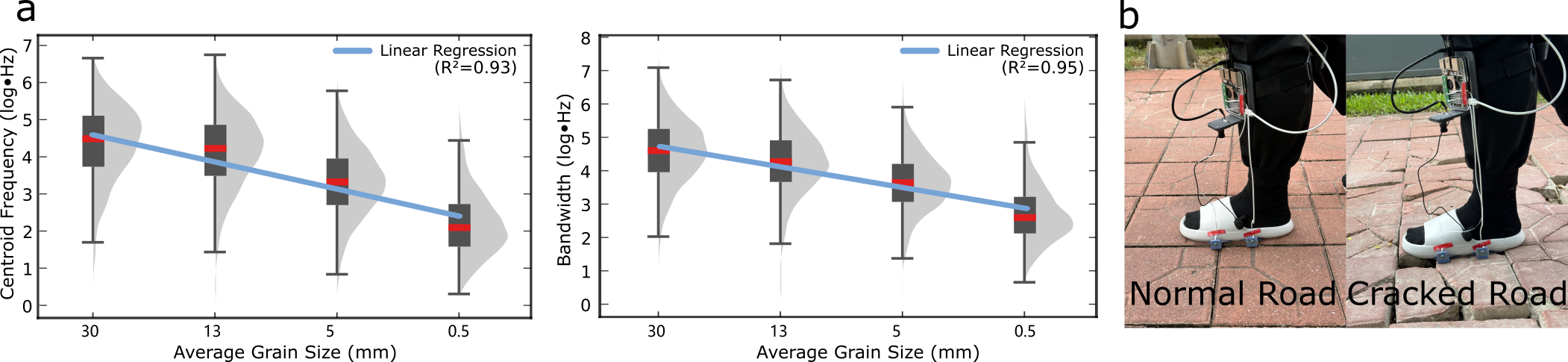}
  \caption{(a) The left plot shows the logarithm of the spectral centroid. The spectral centroid is the center of mass of a sound's spectrum, reflecting its brightness or clarity. The right plot shows the logarithm of the bandwidth. The bandwidth of a spectrum refers to the range of frequencies within which most of the signal's energy is concentrated, typically measured as the difference between the highest and lowest frequencies. (b) The left picture shows the situation of walking on normal ground (slab). The wearable system will detect the ground as a slab. The right picture shows the situation of walking on cracked ground. The wearable system will detect the ground as stone-large. Therefore, we can warn of ground damage by detecting stone-large on a continuous slab road.}
  \Description{(a) The left figure shows the spectral centroid. The spectral centroid is the center of mass of a sound's spectrum, reflecting its brightness or clarity. The right figure shows bandwidth. The bandwidth of a spectrum refers to the range of frequencies within which most of the signal's energy is concentrated, typically measured as the difference between the highest and lowest frequencies. (b) The left picture shows the situation of walking on normal ground (slab). The wearable system will detect the ground as a slab. The right picture shows the situation of walking on cracked ground. The wearable system will detect the ground as stone-large. Therefore, we can warn of ground damage by detecting stone-large on a continuous slab road.}
  \label{fig:grainsize}
\end{figure*}
The vibration data captured by \oursystem contains details of the ground, such as material grain sizes. 
We studied the relationship between the particle diameter of the ground materials and the spectral features of vibration data. We used the data of different grain sizes (including gravel small, gravel mid, gravel large, and sand) from 31 subjects in section~\ref{section:18material} in this study.  The diameters of these four grains are 3\,cm, 1.3\,cm, 5\,mm, and 0.5\,mm. 

We used spectral centroid and bandwidth to characterize the acoustic features of the grains. Specifically, the logarithmic spectral centroid $F_\text{centroid}$ was computed as:
\begin{equation}
    F_\text{centroid} = \log{\sum\left( f\cdot\text{PSD}(f)\right)} - \log{\sum \text{PSD}(f)}
\end{equation}
and the logarithmic bandwidth of concentrated energy $F_\text{bandwidth}$ was computed as:
\begin{equation}
    F_\text{bandwidth} = 0.5\log{\sum \text{PSD}(f) \cdot (f - F_\text{centroid})^2} - 0.5\log{\sum \text{PSD}(f)}
\end{equation}
where $f$ represents the frequencies of the vibration signal spectrum and $\text{PSD}(f)$ represents the power spectral density~\cite{Welch1967}.

We found a strong correlation between the spectral features of vibrations and the grain sizes, specifically, the logarithmic spectral centroid and the logarithmic bandwidth of energy concentration decreased linearly as the grain size decreased (Figure ~\ref{fig:grainsize}). The correlation was confirmed by linear fittings, with an R score of 0.93 and 0.95 for the spectral centroid and bandwidth features, respectively.
This trend is likely due to the lower natural frequencies of smaller particles that are stepped by the shoe.
This finding suggests that frequency domain features can effectively characterize material grain size and may hold potential for broader applications in material property analysis. 

\subsubsection{Application}
The capability of \oursystem to estimate the grain sizes of the ground materials allows it to monitor road conditions, such as identifying cracked road slabs. The benefit is that the classification accuracy for large rocks (Gravel Large) is very high (95$\%$), and the system will classify the damaged slab as Gravel Large. For instance, When a user walks on road pavements, \oursystem can distinguish between normal slabs and cracked ones that move like loose stones. Moreover, it can differentiate the sizes of the cracked slabs, which is helpful for assessing the degree of road damage. \oursystem can potentially assist the maintenance personnels of urban infrastructure in promptly identifying and repairing damaged road pavements, thereby improving the safety and comfort of the urban environment.

\section{Map Construction}
\begin{figure*}[h]
  \centering
  \includegraphics[width=160mm]{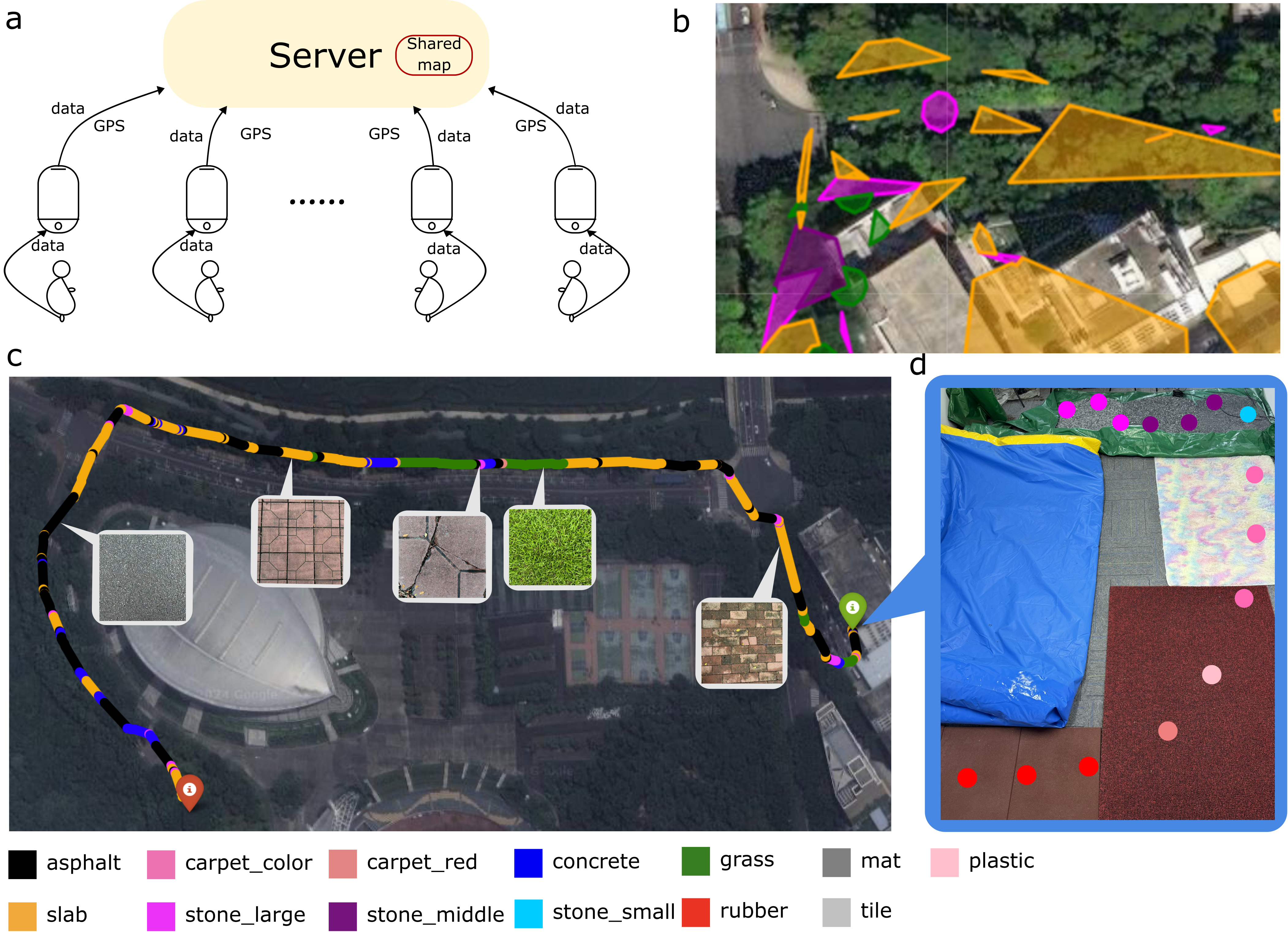}
  \caption{Mapping haptic information for both outdoor and indoor scenes. (a) Collection system structure. Each user can use the Raspberry Pi to collect data and send the data to a mobile phone. The mobile phone sends the data and GPS location to the server. Multiple users can collect data at the same time. (b) Block diagram generated after two users scanned a region collaboratively. Some slab and stone roads are hidden under the trees. (c) Outdoor experiment result: a track of haptic information mapped on a satellite map. (d) Indoor experimental result: different materials were marked by dots of distinct colors, with each color representing a different ground material.}
  \Description{}
  \label{fig:map}
\end{figure*}

\oursystem equipped on pedestrians can create a comprehensive database of a city’s ground information as they walk around. To demonstrate this concept, we conducted an experiment. We utilized vibration sensing and GPS positioning technology to identify and locate different ground materials, which constructed a map of ground information (Figure ~\ref{fig:map}). 

The mapping process includes the following procedures: the mobile phone requests ground information from the Raspberry Pi, the Raspberry Pi sends information to the mobile phone, the mobile phone receives the information and records the current GPS information, the mobile phone packages the ground information and GPS information and sends them to the central server, the server displays the ground material category on the map through the collected ground information and its corresponding GPS signal. It should be noted that since our system is divided into client and server, we can have multiple clients, in other words, we can have multiple pairs of Raspberry Pi and mobile phones to collect data together, which greatly improves the efficiency of collecting data (Figure ~\ref{fig:map}a). To demonstrate the proposed client-server paradigm and data fusion method, we recruited two participants who collected data by wearing \oursystem on their left foot for 20 minutes, and then on their right foot for another 20 minutes. In total, we collected 80 minutes of data. Based on the data collected by \oursystem's clients, we fused the data on the server side and built a graph where the materials were represented by polygons. (Figure ~\ref{fig:map}b).

\oursystem segment the measured vibrations in real time and identify the ground materials for each second.
We evaluated \oursystem on mapping ground information in both outdoor and indoor environment. The user began in an office building and naturally walked over a series of floor materials that were compactly arranged within the room (Figure ~\ref{fig:map}d, zoomed view on the right). This demonstrates the responsiveness of \oursystem in identifying and locating the ground materials. 
The user then walked out of the building and marched on the road of the campus, with road materials majorly consisting of asphalt, stone, grass, concrete, and slab.
Throughout the demonstration, some of the materials are included in the database we used to train the material recognition model, while others are not, such as the slabs that are arranged in different patterns (Figure ~\ref{fig:map}c).

The processed data is utilized to generate a geospatial map centered on the initial coordinate of the trajectory. This map employs a satellite tile base layer, which provides a detailed and realistic backdrop for the visualization. As the trajectory is rendered, it is divided into segments corresponding to the material labels. Each segment is color-coded according to the predefined mapping, with transitions between different materials explicitly indicated by connecting lines.

In addition to the trajectory itself, the map includes markers denoting the start and end points, providing context for the overall path covered. A dynamic legend is also generated, which details the color-material correspondence. This legend is implemented as a draggable element on the map interface, allowing users to reposition it for optimal visibility.

Finally, the complete geospatial visualization is saved as an interactive HTML file. This output allows researchers to examine the material-specific trajectory in a web browser, enabling detailed exploration and potential further analysis. The visualization not only serves as a tool for understanding the spatial distribution of materials and corresponding vibrations produced from walking over those materials but also as a foundation for future work in material classification and environmental interaction analysis based on sensor data. 

\section{Evaluation of Haptics Experiences}
One of the main purposes of recording and mapping haptic information is to create a haptic experience for \gls{xr} users. Each type of ground tactile information can give people a unique tactile feedback experience.
\begin{figure*}[h]
  \centering
  \includegraphics[width=140mm]{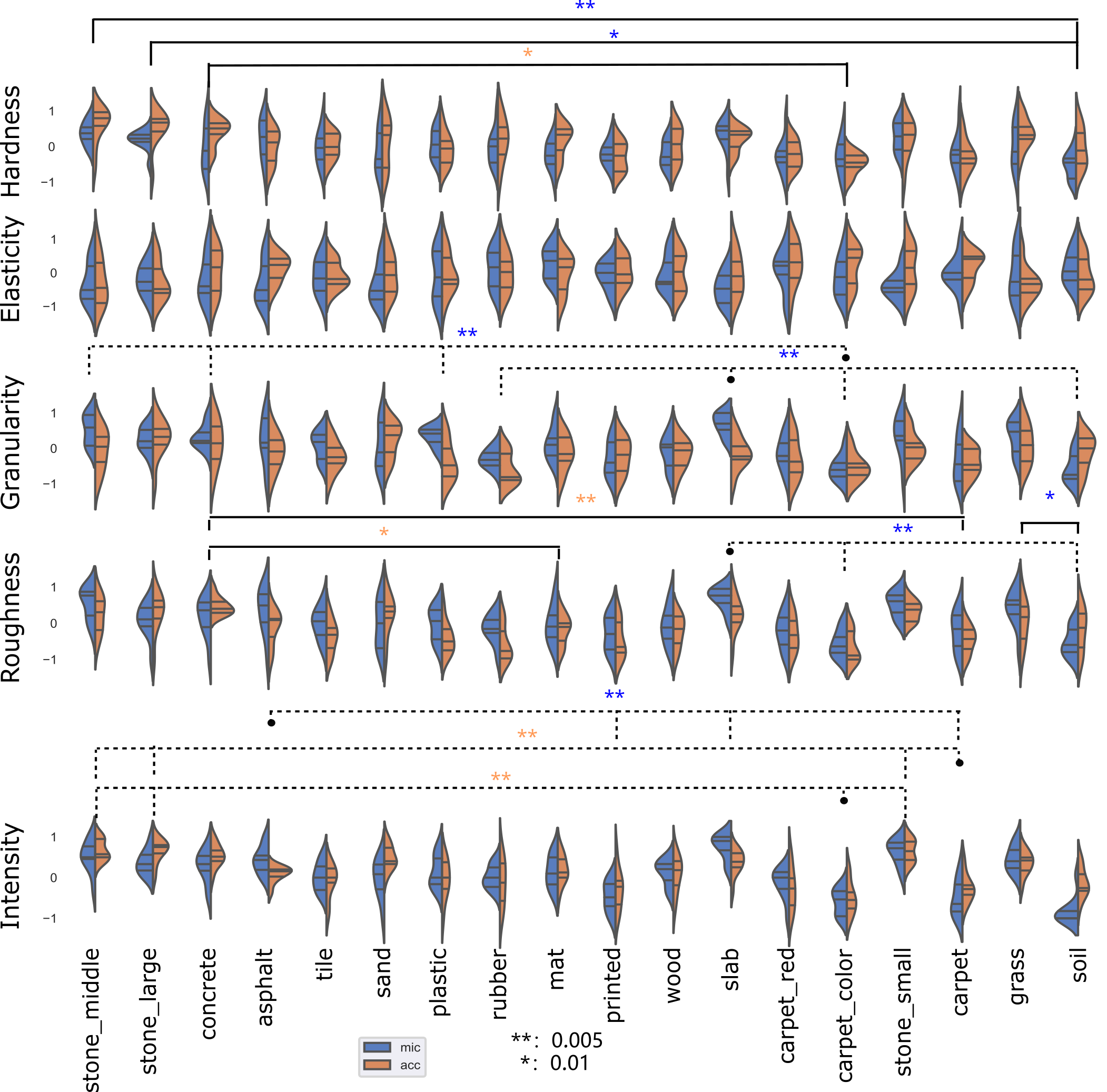}
  \caption{Violin plots of survey data based on users' haptic perceptions of recorded vibrations from the 18 materials, with dotted lines highlighting the differences in subjective ratings between the materials. There are two types of dotted lines. One solid line connects two materials, representing the relationship between the two materials; the other type of line connects multiple materials, including short dashed lines (with one dot) and long dashed lines. There is only one short dashed line, representing material A, and there are several long dashed lines, representing B1, B2... For example, in Intensity, a line connects carpet-color, stone-large, stone-middle, and stone-small. The line connecting carpet-color is a short dashed line with a dot, while the line connecting the other three is a long dashed line, which represents the relationship between carpet color and stone-large, stone-middle, and stone-small. The overall dotted line represents the relationship between A-B1, A-B2. Note: * = p < .01, ** = p < .001.}
  \Description{}
  \label{fig:userstudy}
\end{figure*}
A total of 12 participants were recruited for this study. This study aims to evaluate participants' perception of different tactile characteristics through vibrotactile feedback. This study adopted a within-subject design, and each participant felt all 36 samples, namely the MIC and ACC measurements of the 18 ground textures. Specifically, participants feel vibrations produced by two voice coil motors (QY28R16, 8\,$\Omega$, 2\,W) placed under their right soles, at the heel and ball of the foot, respectively. Both motors are driven by a power amplifier (SUCA AUDIO A900, 300\,W). 
For each trial, the two motors produce vibrations by playing signals recorded from either the MIC or ACC sensor. The sequence in which these recorded signals are presented is completely randomized.
Participants wore noise-canceling headphones playing white noise to block auditory feedback, allowing them to rely solely on haptic feedback. 

During the experiment, participants interacted with a graphical user interface (GUI) that allows them to control and report their subjective perceptions. The interface includes "Play" and "Pause" buttons, enabling participants to control the playback of the recorded signal as needed. After each playback, the participants were instructed to rate their perceptions across five different tactile descriptors: ``Hardness'', ``Elasticity'',``Granularity'', ``Roughness'', and ``Intensity''. Each dimension was rated on a continuum scale from -10 to +10, with +10 representing the strongest tactile sensation represented by the tactile descriptor and -10 indicating the complete absence of that sensation.
Participants, unaware of the specific values between -10 and 10, used sliders to select values based on their subjective standards, quantifying their feelings about the vibration characteristics of the current feedback. The GUI design ensured that participants received no information about the recording source (MIC or ACC), allowing them to focus solely on their haptic experience. Each participant experienced all 36 samples only once. Although no repeated trials were performed, we randomized the trial order to reduce the influence of order effects.

\begin{table*}[h]
\caption{Surveys obtain users' evaluation of ground feedback from five dimensions: ``Hardness'', ``Elasticity'', ``Granularity'', ``Roughness'', and ``Intensity''. We analyzed their difference with \gls{anova} and show the p-value and the Bonferroni-adjusted p-value.}
\begin{tabular}{@{}c|ccccc@{}}
\toprule
MIC/ACC    & Hardness & Elasticity & Granularity & Roughness & Intensity \\ \midrule
p-value    & 0.0018/6.18e-06        & 0.28/0.71          & 5.8e-10/0.0049           & 4.7e-9/5.5e-9         & 1.5e-24/1.2e-16         \\
Bonferroni & 0.0183/6.1e-5        & 1/1          & 5.8e-9/0.0393           & 4.7e-8/5.5e-8         & 1.5e-23/1.3e-15         \\ \bottomrule
\end{tabular}
\label{tab:userstudy}
\end{table*}

For each MIC or ACC recording, we assessed the differences in users' subjective ratings across the five perceptual dimensions of haptic feedback using repeated measures \gls{anova} (Table~\ref{tab:userstudy}). We corrected the reported p-values using Bonferroni correction. We also identified pairs of materials with significant perceptual differences using Student's t-test, with results summarized in Figure~\ref{fig:userstudy}, where the five subjective dimensions were compared independently. 
We highlight a few key points worth noting. 
There is a significant difference in perception between soft and hard materials. Specifically, the subjects' perception of soft materials and their corresponding rating are more concentrated.
The subjects are able to distinguish between soft and hard materials. For example, in the comparison between carpet-color and three types of stones with different particles using t-test, the p values are all less than 0.005, which is a statistically significant difference.
Carpet-color is very different from plastic, concrete and stone-middle. One participant commented: ``I can't feel any Granularity from this feedback signal (carpet-color)''. In addition, there are also huge differences between slab, rubber, carpet-color, and soil.
There is a huge difference in roughness between carpet and concrete, and between slab and soft surfaces, such as carpet-color and soil.
We also observed significant difference between soft ground, carpet-color and carpet. Stones of different sizes that have a strong granular feel. 
Elasticity showed no much difference, but possibly be due to the hardware limitation of our haptic feedback device utilizing two voice coil actuators.

\section{Limitations and Future Work}
\paragraph{Decoding Ground Information}
The \oursystem model proposed in this work can discriminate among 18 different materials stepped on by users. However, it encounters difficulties in accurately identifying certain materials, such as asphalt, slab, and concrete. These errors become more pronounced in cross-user studies, indicating a need for improved generalizability of the \oursystem model.
We plan to further enhance the classifier’s performance to address user variability in \oursystem. To enable large-scale application of this technology in real-world scenarios, we will explore data augmentation techniques, such as incorporating more users with unique gait behaviors. Moreover, we will expand the scope of data collection to include more categroies of gournd materials and even composite material surfaces.
In addition, the fundamental physics of foot-ground interaction can be better understand via experiment comprehensively evaluate the impact of different types of materials on vibrations produced during walking. With a better physical modelling, vibration data of a broad range of ground materials can be simulated and supplement the limited real-world data, to enhance the training of the \oursystem model. In addition, to address the noise interference issue of the MIC, we plan to first accurately obtain a batch of cleaned sounds, and then before putting them into the model, we actively add larger Gaussian noise to the signal to improve the robustness of our model to noise.

\paragraph{Scalability and Edge Computing}
Moreover, future research will focus on data collection and processing on edge devices. We will design lightweight model architectures and optimize computational resource requirements to efficiently run on resource-constrained edge devices. We will explore distributed computing or edge collaborative learning methods to distribute computational loads across multiple devices, further enhancing the system's scalability and practicality.
As for the data, this experiment sets 18 ground materials, which means we can add more new materials in future work. However, among the newly added materials, there could be materials with properties very similar to the existing 18 materials. Therefore, we need to improve the selection of materials and the model's ability to decode the properties of materials.

\paragraph{Immersive Haptic Feedback}
We plan to utilize the collected vibration data to provide vibrotactile feedback, aiding in the development of haptic feedback-enhanced reality shoes. These advanced shoes can simulate ground sensations that do not exist in real life, offering an immersive virtual reality experience. Through this approach, we aim to further expand the application of shoe technology in haptic feedback and virtual reality.

\paragraph{Wearability and Convenience}
Although the transmission plate enhances the signal, it affects running or strenuous exercise of the lower limbs. We plan to learn from Adidas's way of installing sensors on the insoles~\cite{adidas2024GMR} and set our sensors as wireless rechargeable devices and place them on the side of the shoe or in the midsole. In addition, we also need to improve the model to enhance its recognition ability when there is no transmission plate.

\paragraph{Decoding Human Activities}
We also plan to extend the current classification framework to human activity recognition tasks. By integrating shoe information, we aim to identify complex activity patterns. We will collect and annotate more data involving various types of activities and use deep learning models to capture the relationship between shoe characteristics and activity types, significantly broadening the applicability of our research.

\paragraph{Joint Mapping with Vehicles}
Although vehicles cannot access certain rugged roads to collect data, we can achieve a way for humans and vehicles to collect data together. In some larger areas, vehicles are used to collect and identify ground data~\cite{ramos2021shifts}; on rugged mountain roads or in areas where vehicles are prohibited from entering, humans are used to collect and identify data. Therefore, we also need to build a system that can collect and identify vehicles and humans at the same time, because the models of vehicle identification and human identification of the ground and the form of data required may be somewhat different.

\section{Conclusion}
We present \oursystem, a wearable system that can obtain ground details and pedestrian haptic information. We explored the use of vibrations to identify the material of the ground. We fused the temporal and spectral features of wideband vibrations and proposed a deep learning model based on ResNet framework, which can identify ground materials with 96.1$\%$ accuracy. In addition, we tested \oursystem on identifying dry and wet ground surfaces and obtained 96.9$\%$ accuracy. We conducted a statistical analysis of materials with different particle sizes and found that their sizes have a strong correlation with log spectrum centroid and log bandwidth of the vibration data captured by \oursystem, which allow \oursystem to detect road damages. Finally, We demonstrate the capability of \oursystem in building a map of haptic information. Via a user study, we also validated the effectiveness of providing a haptic experience based on the recordings of \oursystem. The results showed that users could identify material properties solely by feeling the vibrations produced on their soles.

\begin{acks}
This work was supported by the National Natural Science Foundation of China under Grant BA2440700124, by the Shenzhen High Level Talent under Grant ZX20230587, Shenzhen Science and Technology Innovation Program under Grant GXWD20231128112815001, and the National Research Foundation of Korea (NRF) grant funded by the Korea government (MSIT) (RS-2023-00210001).
\end{acks}

\bibliographystyle{ACM-Reference-Format}
\bibliography{refs.bib}

\end{document}